\documentclass[prl,twocolumn,superscriptaddress,fixfloat]{revtex4}
\usepackage{amsmath}
\usepackage{amssymb}
\usepackage{graphicx}
\begin{document}

\title{A neutron scattering study of the interplay between structure and magnetism in Ba(Fe$_{1-x}$Co$_{x}$)$_2$As$_2$}

\author{C. Lester}
\affiliation{H.H. Wills Physics Laboratory, University of Bristol, Tyndall Ave., Bristol,
BS8 1TL, UK}

\author{Jiun-Haw Chu}
\affiliation{Geballe Laboratory for Advanced Materials and Department of Applied Physics, Stanford University, Stanford, CA 94305}
\affiliation{Stanford Institute for Materials and Energy Sciences, SLAC National Accelerator Laboratory, 2575 Sand Hill Road, Menlo Park, CA 94025}

\author{J. G. Analytis}
\affiliation{Geballe Laboratory for Advanced Materials and Department of Applied Physics, Stanford University, Stanford, CA 94305}
\affiliation{Stanford Institute for Materials and Energy Sciences, SLAC National Accelerator Laboratory, 2575 Sand Hill Road, Menlo Park, CA 94025}

\author{S. Capelli}
\affiliation{Institut Laue-Langevin, 38042 Grenoble, France}

\author{A. S. Erickson}
\affiliation{Geballe Laboratory for Advanced Materials and Department of Applied Physics, Stanford University, Stanford, CA 94305}
\affiliation{Stanford Institute for Materials and Energy Sciences, SLAC National Accelerator Laboratory, 2575 Sand Hill Road, Menlo Park, CA 94025}

\author{C. L. Condron}
\affiliation{Stanford Syncrhrontron Radiation Laboratory, SLAC National Accelerator Laboratory, 2575 Sand Hill Road, Menlo Park, CA 94025}

\author{M. F. Toney}
\affiliation{Stanford Syncrhrontron Radiation Laboratory, SLAC National Accelerator Laboratory, 2575 Sand Hill Road, Menlo Park, CA 94025}

\author{I. R. Fisher}
\affiliation{Geballe Laboratory for Advanced Materials and Department of Applied Physics, Stanford University, Stanford, CA 94305}
\affiliation{Stanford Institute for Materials and Energy Sciences, SLAC National Accelerator Laboratory, 2575 Sand Hill Road, Menlo Park, CA 94025}

\author{S.M. Hayden}
\affiliation{H.H. Wills Physics Laboratory, University of Bristol, Tyndall Ave., Bristol,
BS8 1TL, UK}

\begin{abstract}
 Single crystal neutron diffraction is used to investigate the magnetic and structural phase diagram of the electron doped superconductor Ba(Fe$_{1-x}$Co$_x$)$_2$As$_2$. Heat capacity and resistivity measurements have demonstrated that Co doping this system splits the combined antiferromagnetic and structural transition present in BaFe$_2$As$_2$ into two distinct transitions. For $x$=0.025, we find that the upper transition is between the high-temperature tetragonal and low-temperature orthorhombic structures with ($T_{\mathrm{TO}}=99 \pm 0.5$~K) and the antiferromagnetic transition occurs at $T_{\mathrm{AF}}=93 \pm 0.5$~K. We find that doping rapidly suppresses the antiferromagnetism, with antiferromagnetic order disappearing at $x \approx 0.055$.  However, there is a region of co-existence of antiferromagnetism and superconductivity. The effect of the antiferromagnetic transition can be seen in the temperature dependence of the structural Bragg peaks from both neutron scattering and x-ray diffraction.  We infer from this that there is strong coupling between the antiferromagnetism and the crystal lattice.
\end{abstract}


\maketitle

\section{Introduction}
The discovery of the Fe-pnictide superconductors \cite{Kamihara2008a} offers a new opportunity to investigate superconductivity and its mechanism in a new class of materials. The pnictide superconductors share some features of the high-$T_c$ cuprate superconductors: the superconductivity occurs in the proximity of an antiferromagnetic compound and can be induced by doping. However, there are also significant differences, for example, the parent antiferromagnets are metallic\cite{analytis2009a}. The two most widely studied systems are those based on RFeAsO$_{x}$F$_{y}$ (with R=Nd,Sm,Pr,La) \cite{Kamihara2008a,Chen2008a,Ren2008a} and AFe$_2$As$_2$ (A=Ca,Sr,Ba) \cite{Rotter2008a,Sefat2008a,Ni2008a,Chu2009a,Ning2009a}, known as ``1111'' and ``122'' respectively.  The 122 system can be doped at the A \cite{Rotter2008b} or Fe \cite{Sefat2008a,Ni2008a,Chu2009a,Ning2009a} sites to achieve superconductivity.

The determination of the phase diagram as a function of doping in the Fe-pnictide systems is an important step to understanding the superconductivity. However, there appears to be significant differences between the various systems. The general tendency is for doping to suppress orthorhombic structure and antiferromagnetic order \cite{delaCruz2008a,Huang2008_LaFeAs} present at low temperatures and for superconductivity to emerge. However, the manner in which this transition occurs appears to vary between systems.  In CeFeAsO$_{1-x}$F$_{x}$, the magnetic order is suppressed before superconductivity develops and superconductivity exists both in the orthorhombic and tetragonal structures \cite{Zhao2008a}.  Whereas in LaFeAsO$_{1-x}$F$_{x}$, the orthorhombicity and magnetic order disappear abruptly \cite{Huang2008_LaFeAs,Luetkens2009a} just as the superconductivity develops. In this paper, we use neutron scattering to investigate the magnetic and structural phase diagram of the electron doped superconductor Ba(Fe$_{1-x}$Co$_x$)$_2$As$_2$ \cite{Sefat2008a,Ni2008a,Chu2009a,Ning2009a}. This material has risen to prominence recently because high-quality single crystals can be prepared which show two anomalies in the heat capacity $C(T)$ and resistivity $\rho(T)$ for small Co doping \cite{Ni2008a,Chu2009a}. Here we use neutron and x-ray scattering to identify the nature of the transitions and determine the phase diagram. In contrast to observations in LaFeAsO$_{1-x}$F$_{x}$ and CeFeAsO$_{1-x}$F$_{x}$, we find that the orthorhombic transition and magnetic order persist to higher dopings and coexist with the superconductivity.

\section{Experiment}
Single crystals of Ba(Fe$_{1-x}$Co$_x$)$_2$As$_2$ with $x$=0.025, 0.045 and 0.051 were grown by a self flux method \cite{Chu2009a}. The crystals were plate-like with masses up to 15~mg and dimensions up to $4 \times 3 \times 0.3$~mm. Crystals from the same batches have been characterized by resistivity, heat capacity, Hall effect and susceptibility \cite{Chu2009a}.  Neutron diffraction measurements were made on three crystals. Neutron Laue measurements showed these crystals to be single grain. The single crystal neutron diffraction data presented in this paper were collected using the D10 four-circle diffractometer at the Institut Laue-Langevin. We used an incident wavelength of $\lambda$=2.364~\AA\ and a 2-D area detector. A graphite filter was used to reduce $\lambda/2$ contamination in the incident beam. BaFe$_2$As$_2$ undergoes a tetragonal(T)-orthorhombic(O) ($I4/mmm \rightarrow Fmmm$) structural transition at 134~K \cite{Rotter2008a}. The low temperature orthorhombic phase is described by the $Fmmm$ space group with $b<a<c$.  Because of the small difference in the $a$ and $b$ lattice parameters for the doped Ba(Fe$_{1-x}$Co$_x$)$_2$As$_2$ samples studied here, we do not directly resolve Bragg peaks from different orthorhombic domains in the present experiment.  We therefore use the tetragonal $I4/mmm$ space group to label reciprocal space except where stated. The transformation between the two descriptions for $T \geq T_{\mathrm{TO}}$  is
$h_{\mathrm{O}}=h_{\mathrm{T}}+k_{\mathrm{T}}$, $k_{\mathrm{O}}=h_{\mathrm{T}}-k_{\mathrm{T}}$,
$l_{\mathrm{O}}=l_{\mathrm{T}}$.  The tetragonal lattice parameters at $T=200$~K were: $a$=3.959~\AA,  $c$=12.97~\AA\ ($x$=0.025); $a$=3.955~\AA, $c$=12.95~\AA\ ($x$=0.045); $a$=3.955~\AA, $c$=12.95~\AA\ ($x$=0.051). High resolution x-ray diffraction has also been performed on beamline 2-1 at Stanford Synchrontron Radiation Laboratory, using an incident x-ray energy of 11.7 keV. Samples are powdered from single crystals of the same compositions.

\begin{figure}
\begin{center}
\includegraphics[width=0.8\linewidth]{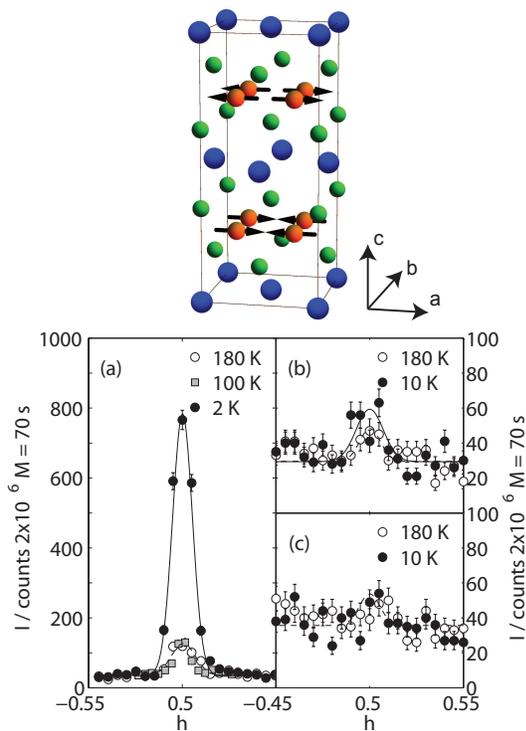}
\end{center}
\caption{(color online) The top panel shows the antiferromagnetic structure of BaFe$_2$As$_2$ \cite{Huang2008_BaFe2As2,Su2009a,Kofu2009a}. Axes refer to orthorhombic notation. (a)-(c) $\mathbf{Q}$ scans through the (1/2 1/2 1)$_{\mathrm{T}}$ peak for various values of Co doping $x$ in Ba(Fe$_{1-x}$Co$_x$)$_2$As$_2$.  The concentrations are $x$=0.025, 0.045, 0.051 for (a), (b) and (c) respectively.  The smaller peaks in (a) present for $T$=180~K and $T$=100~K are due to is due to unfiltered $\lambda/2$ neutron scattering from the (1 1 2) nuclear Bragg. The increased scattering observed at 2~K is due to magnetic order. The samples used in (b) and (c) are smaller than in (a).  The plots have been appropriately scaled to compensate.  Increased scattering at low temperatures is also observed for $x$=0.045.}
\label{Fig:AF_peaks}
\end{figure}

\begin{figure}
\begin{center}
\includegraphics[width=0.98\linewidth]{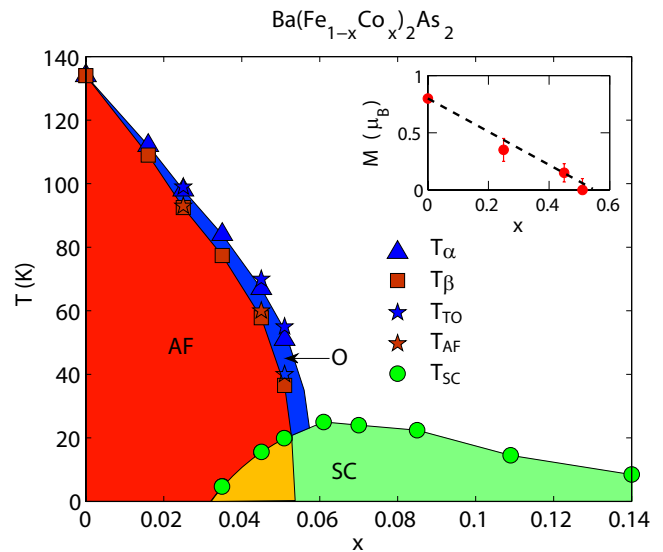}
\end{center}
\caption{(color online) The phase diagram for Ba(Fe$_{1-x}$Co$_x$)$_2$As$_2$ determined from neutron scattering, $C(T)$ and $\rho(T)$ \cite{Chu2009a}. The TO structural transition is clearly observed by neutron scattering, heat capacity and resistivity. Neutron scattering shows a magnetic Bragg peak for x=0 \cite{Huang2008_BaFe2As2,Su2009a,Kofu2009a} ,0.025, 0.051.  The inset shows the doping dependence of the ordered based on the present work and Refs.~\cite{Huang2008_BaFe2As2,Su2009a,Kofu2009a}. Phase labels are O(orthorhombic), T(tetragonal), SC(superconducting), AF(antiferromagnetic).}
\label{Fig:phase_diagram}
\end{figure}

\begin{figure}
\begin{center}
\includegraphics[width=0.98\linewidth]{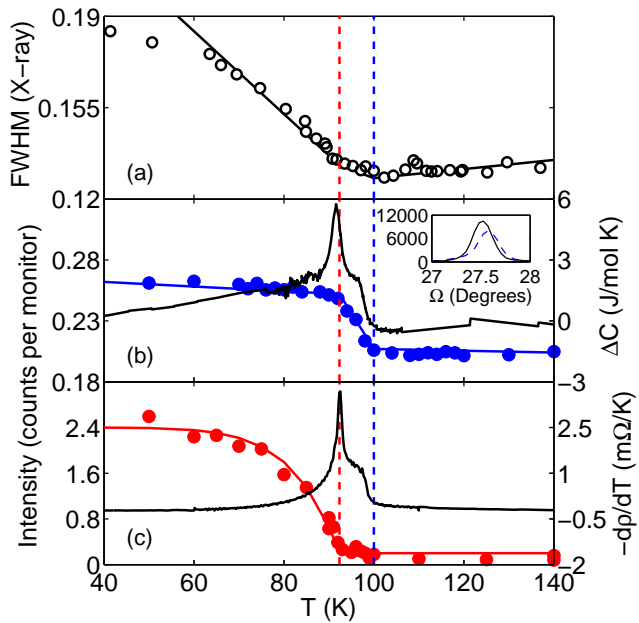}
\end{center}
\caption{(Color online) Identification of the phase transitions in Ba(Fe$_{1-x}$Co$_x$)$_2$As$_2$ (x=0.025). (a) $T$-dependence of the width of the (112) powder line determined from x-ray diffraction.  (b) $T$-dependence of integrated intensity (blue circles) of the (112)$_{\mathrm{T}}$ nuclear peak determined from $\omega-2 \theta$ scans. Inset shows the Bragg profile $T$=10~K and 120~K. The solid back line is the heat capacity $C(T)$ from Ref.~\cite{Chu2009a}. The magnetic structure of BaFe$_2$As$_2$ \cite{Huang2008_BaFe2As2,Su2009a,Kofu2009a} is shown at the top of the diagram. (c) $T$-dependence of the integrated intensity (red circles) of the magnetic (1/2 1/2 1)$_{\mathrm{T}}$ Bragg peak. Solid line is $-d\rho/dT$ \cite{Chu2009a}. }
\label{Fig:Peak_Int_width_25}
\end{figure}

\begin{figure}
\begin{center}
\includegraphics[width=0.98\linewidth]{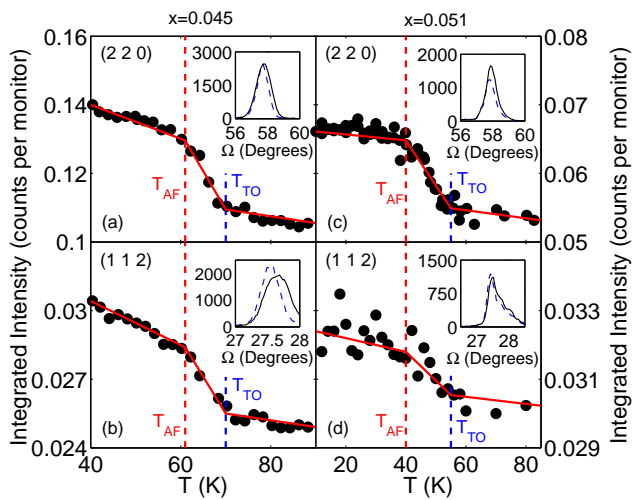}
\end{center}
\caption{$T$-dependence of the integrated intensity of (220) and (112) nuclear Bragg peaks for the more highly doped $x=0.045$ and $x=0.051$ samples.  The kinks appear to correspond to tetragonal-orthorhombic (TO) and antiferromagnetic (AF) transitions.  The insets show example Bragg profiles for $T>T_{\mathrm{TO}}$ (dotted line) and $T<T_{\mathrm{AF}}$ (solid line).}
\label{Fig:Peak_Int_45_51}
\end{figure}

\section{Results}
Previous studies of specific heat and neutron scattering on BaFe$_2$As$_2$ have shown that this parent compound shows a single magnetic/structural phase transition at $T \approx $134~K \cite{Rotter2008a,Huang2008_BaFe2As2,Su2009a,Wilson2009a}.  The structural transition corresponds to a tetragonal-orthorhombic (TO) ($I4/mmm \rightarrow Fmmm$) transformation \cite{Rotter2008a} and the magnetic transition is to a antiferromagnetic (AF) ordered state \cite{Huang2008_BaFe2As2,Su2009a,Kofu2009a} with magnetic moments pointing along the orthorhombic $a$ axis and aligned antiferromagnetically along the $a$ and $c$ axes, and ferromagnetically along the $b$ axis \cite{Huang2008_BaFe2As2,Su2009a,Kofu2009a} (see Fig.~\ref{Fig:AF_peaks}).
In contrast to BaFe$_2$As$_2$, the parent compound of the 1111 series, LaFeAsO, shows two transitions with decreasing temperature. The structural TO transition is followed by the AF transition. Interestingly, Co-substitution on the Fe sites splits the single transition in BaFe$_2$As$_2$ into two transitions which can be tracked across the Ba(Fe$_{1-x}$Co$_x$)$_2$As$_2$ phase diagram using heat capacity [$C(T)$] and resistivity [$\rho(T)$] measurements \cite{Ni2008a,Chu2009a}. The positions of the anomalies determined from heat capacity and resistivity \cite{Chu2009a} are shown in Fig.~\ref{Fig:phase_diagram}.

In order to identify the nature of the phases delineated by $C(T)$ and $\rho(T)$, we investigated three different compositions of Ba(Fe$_{1-x}$Co$_x$)$_2$As$_2$. The magnetic structure established in Refs.~\cite{Huang2008_BaFe2As2,Su2009a,Kofu2009a} gives rise to magnetic reflections at positions such that the orthorhombic indices are $h_{\mathrm{O}}$=odd, $k_{\mathrm{O}}$=even and $l_{\mathrm{O}}$=odd.   We measured four inequivalent magnetic reflections. Fig.~\ref{Fig:AF_peaks}(a)-(c) shows scans through the (1/2,1/2,1)$_{\mathrm{T}}$ position, which corresponds to the (1 0 1)$_\mathrm{O}$ magnetic reflection, for the three different cobalt dopings. Fig.~\ref{Fig:AF_peaks}(a) shows scans through the (1/2 1/2 1) position for the $x$=0.025 sample with $T$=10~K, 100~K, 180~K.
Our data is slightly contaminated by unfiltered $\lambda/2$ neutrons which scatter from the (1 1 2) nuclear bragg peak. This scattering produces a small temperature-independent peak at the same spectrometer position ($\omega,2 \theta$) as the (1/2 1/2 1) magnetic peak. The peak can be seen for $T$=100~K and 180~K. At low temperatures ($T=2~K$), we observe additional scattering [see Fig.~\ref{Fig:AF_peaks}(a)] due to antiferromagnetic order.  Our limited dataset is consistent with the magnetic structure reported in Ref.~\cite{Huang2008_BaFe2As2,Su2009a,Kofu2009a} and we do not observe incommensurate peaks as suggested by recent NMR measurements \cite{Ning2009_IC}. Fig.~\ref{Fig:Peak_Int_width_25}(c) shows the temperature dependence of the integrated intensity of the (1/2,1/2,1) peak. Fitting a mean field form to the order parameter near the antiferromagnetic transition temperature $T_{\mathrm{AF}}$, we obtain $T_{\mathrm{AF}}=93 \pm 0.5$. Thus the lower $T_{\beta}$ transition previously identified \cite{Chu2009a} in $C(T)$ and $\rho(T)$ [reproduced in Fig.~\ref{Fig:Peak_Int_width_25}] is the antiferromagnetic transition.  An estimate of the low-temperature ordered moment can be obtained by normalizing the data to nuclear Bragg reflections. Using 27 reflections and assuming the magnetic structure determined in Ref.~\cite{Huang2008_BaFe2As2,Su2009a,Kofu2009a}, we determine the ordered moment to be $m=0.35 \pm 0.1~\mu_{\mathrm{B}}$.

Our neutron data can also be used to identify the temperature ($T_{\mathrm{TO}}$) at which tetragonal-orthorhombic (TO) structural transition occurs.  The small difference between the $a$ and $b$ lattice parameters in the orthorhombic phase means that we are unable to resolve the (202)$_{\mathrm{O}}$ and (022)$_{\mathrm{O}}$ peaks due to the different orthorhombic domains (for example, as in Ref.~\cite{Goldman2008a}). However, the TO phase transition can be detected through the temperature dependence of the intensity, $I(T)$, and width, $\Gamma(T)$, of the (112)$_{\mathrm{T}}$ and (220)$_{\mathrm{T}}$ Bragg peaks. The Bragg peaks intensities are not directly proportional to the order parameter of the TO transition because of formation of domains and extinction, however, they are nevertheless a probe of structural changes.  Fig.~\ref{Fig:Peak_Int_width_25}(b) shows the temperature dependence of the integrated intensity of the (112)$_{\mathrm{T}}$ Bragg peak measured by neutron diffraction.   The previously identified $T_{\alpha}$ transition coincides with the onset of a rapid intensity increase and is therefore identified with the TO structural phase transition. We also observed a broadening in the (112) peak using x-ray diffraction on a powdered sample of the same composition. This is shown in Fig.~\ref{Fig:Peak_Int_width_25}(a). In addition to the kink at the TO transition, we observe a second kink at $T_{\mathrm{AF}}$. This kink is observed both in the neutron and the x-ray measurements. The existence of the second kink suggests that the antiferromagnetic transition affects the crystal lattice. Two possible mechanisms for the change in neutron intensity are: (i) a sudden change in the first temperature derivative of the order parameter associated the structural transition at $T_{\mathrm{AF}}$; (ii) a change in the orthorhombic domain structure causing extinction release.  Both mechanisms require coupling between the antiferromagnetic order and the lattice.

We also investigated samples with $x$=0.045 and $x$=0.051. Fig.~\ref{Fig:AF_peaks}(b) shows that for $x$=0.045 there is a small increase in scattering on lowering the temperature from 180~K to 10~K. However, for $x$=0.051 no increase in scattering is observed on lowering the temperature from 180~K to 10~K. The $T_{\alpha}$ and $T_{\beta}$ anomalies in $\rho(T)$ persist for the $x$=0.045 and $x$=0.051 samples.  Thus, we conclude that both compositions order magnetically but the ordered moment of the $x=0.051$ sample is probably below the threshold for detection of the present experiment.  As with the $x$=0.025 data, we estimate the ordered moment for each doping by normalizing the data to a set of nuclear Bragg reflections and assuming the magnetic structure of BaFe$_2$As$_2$ \cite{Huang2008_BaFe2As2,Su2009a,Kofu2009a}.  The results are shown in the inset to Fig.~\ref{Fig:phase_diagram}.   In addition to studying the magnetic scattering from the more highly doped $x$=0.045 and 0.051 samples,  we also investigated the (112) and (220) nuclear Bragg peaks.  Fig.~\ref{Fig:Peak_Int_45_51} shows the temperature dependence of the integrated intensity $I(T)$.  Just as for the $x=0.025$ sample, we observe kinks in the $I(T)$ curves at the $T_{\alpha}$ and $T_{\beta}$ anomalies identified from $\rho(T)$. Showing again that the antiferromagnetism is strongly coupled to the lattice.

\section{Discussion}

In Fig.~\ref{Fig:phase_diagram} we collect together the positions of the anomalies determined by neutron scattering, heat capacity and resistivity \cite{Chu2009a}.  The inset shows the doping dependence of the order moment from this work ($x\geq 0.025$) and Refs.~\cite{Huang2008_BaFe2As2,Su2009a,Kofu2009a} for $x$=0.  For $x$=0, is well known that BaFe$_2$As$_2$ shows a combined antiferromagnetic/structural transition \cite{Rotter2008a,Huang2008_BaFe2As2,Su2009a,Wilson2009a} which is believed to be second order or weakly first order. On doping this transition is split into two separate anomalies \cite{Ni2008a,Chu2009a} which can be seen in heat capacity and resistivity. Here we have identified the upper anomaly as the TO structure transition and the lower one as antiferromagnetism. The inset to Fig.~\ref{Fig:phase_diagram} shows how the ordered antiferromagnetic moment is suppressed with doping. The present data are consistent with the moment disappearing at the same doping as the $\beta$ transition and at approximately $x$=0.055. It is interesting to compare the present phase diagram to the other iron pnictides. In the related 122 material Ba$_{1-x}$K$_{x}$Fe$_{2}$As$_{2}$,  doping also suppresses the TO and AF transitions and a region of co-existence of superconductivity and antiferromagnetism is observed \cite{Rotter2008b,Chen2009a}. Such behavior has also been observed in the 1111 system SmFeAsO$_{1-x}$F$_x$ \cite{Drew2009a}. In contrast, the other 1111 systems LaFeAsO$_{1-x}$F$_x$ \cite{Huang2008_LaFeAs,Luetkens2009a} and CeFeAsO$_{1-x}$F$_x$ \cite{Zhao2008a} do not show a coexistence of antiferromagnetism and superconductivity. Thus, there appear to be significant differences between the different FeAs superconductor systems.

It has been argued \cite{Yildirim2008a} that the magnetic interactions in the undistorted high-temperature tetragonal phase of the Fe pnictides are highly frustrated. Furthermore, the system can relieve this frustration by making an orthorhombic distortion of the lattice \cite{Yildirim2008a}. This means there is strong coupling coupling between the antiferromagnetism and the lattice. Whilst the structural distortion and antiferromagnetic transitions occur simultaneously for BaFe$_2$As$_2$ and sister parent compounds, $T_{\mathrm{TO}}>T_{\mathrm{AF}}$ in the 1111 compounds and Co doped BaFe$_2$As$_2$ \cite{Ni2008a}. It has been suggested that the splitting \cite{Fang2008a,Mazin2009a} is due to the formation of fluctuating antiferromagnetic domains below $T_{\mathrm{TO}}$ which become pinned at the lower transition $T_{\mathrm{AF}}$ or an Ising transition at $T_{\mathrm{TO}}$ followed by an antiferromagnetic transition at $T_{\mathrm{AF}}$ \cite{Qi2009a}.  Both these scenarios involve coupling to the lattice and might explain the two kinks in the $I(T)$ curves.

During the preparation of this manuscript, we became aware of a neutron scattering study by Pratt \textit{et al.} \cite{Pratt2009a} on a sample of composition Ba(Fe$_{1-x}$Co$_x$)$_2$As$_2$ ($x$=0.047). The results are in agreement with those presented here for the composition $x$=0.045. The authors associate the lower transition with antiferromagnetism and the upper with the TO structural phase transition.

\section{Conclusion}

In summary, in order to elucidate the phase diagram of Ba(Fe$_{1-x}$Co$_x$)$_2$As$_2$, we have performed neutron and x-ray diffraction studies on three samples with different Co concentrations. Magnetic neutron diffraction is used to identify the upper anomaly ($T_{\alpha}$) seen in heat capacity, resistivity and other bulk measurements \cite{Ni2008a,Chu2009a} as the tetragonal-orthorhombic structural phase transition and the lower ($T_{\beta}$) anomaly as the antiferromagnetic (spin density wave) transition. We find that doping rapidly suppresses the antiferromagnetism, with antiferromagnetic order disappearing at $x \approx 0.055$.  However, there is a region of co-existence for antiferromagnetism and superconductivity. Measurements of the $T$-dependence of the intensity and width of the structural Bragg peaks show that coupling to the lattice plays an important part in the antiferromagnetic transition.

\section{Acknowledgements}
Work at Bristol is supported by the EPSRC. Work at Stanford is supported by the Department of Energy, Office of Basic Energy Sciences, Division of Materials Sciences and Engineering, under contract DE-AC02-76SF00515. A portion of this work was carried out at the Stanford Synchrotron Radiation Laboratory, a national user facility operated by Stanford University on behalf of the U.S. Department of Energy, Office of Basic Energy Sciences.

%

\end{document}